\documentclass[conference]{IEEEtran}
\IEEEoverridecommandlockouts

\usepackage{cite}
\usepackage{amsmath,amssymb,amsfonts}
\usepackage{algorithmic}
\usepackage{graphicx}
\usepackage{textcomp}
\usepackage{xcolor}
\usepackage{amsmath} 
\usepackage{booktabs} 
\usepackage{multirow} 
\usepackage{array} 
\usepackage{float}
\usepackage{placeins}
\usepackage{hyperref}
\usepackage{caption}

\usepackage[belowskip=0pt,aboveskip=3pt]{caption}

\def\BibTeX{{\rm B\kern-.05em{\sc i\kern-.025em b}\kern-.08em
    T\kern-.1667em\lower.7ex\hbox{E}\kern-.125emX}}
\begin{document}

\title{CADRE: Customizable Assurance of Data Readiness in Privacy-Preserving Federated Learning}


\author{\IEEEauthorblockN{
Kaveen Hiniduma\IEEEauthorrefmark{1}\IEEEauthorrefmark{2},
Zilinghan Li\IEEEauthorrefmark{2},
Aditya Sinha\IEEEauthorrefmark{2}\IEEEauthorrefmark{3},
Ravi Madduri\IEEEauthorrefmark{2},
Suren Byna\IEEEauthorrefmark{1}
}
\IEEEauthorblockA{
\IEEEauthorrefmark{1}The Ohio State University
\IEEEauthorrefmark{2}Argonne National Laboratory 
\IEEEauthorrefmark{3}University of Illinois Urbana-Champaign
}
\IEEEauthorblockA{
\{hiniduma.1, byna.1\}@osu.edu, \{zilinghan.li, madduri\}@anl.gov, aditya47@illinois.edu
}
}

\maketitle

\begin{abstract}
Privacy-Preserving Federated Learning (PPFL) is a decentralized machine learning approach where multiple clients train a model collaboratively. PPFL preserves the privacy and security of a client's data without 
 exchanging it.
However, ensuring that data at each client is of high quality and ready for federated learning (FL) is a challenge due to restricted data access.
In this paper, 
we introduce 
CADRE (\underline{C}ustomizable \underline{A}ssurance of \underline{D}ata \underline{RE}adiness) for federated learning (FL), a novel framework that allows users to define custom data readiness (DR) metrics, rules, and remedies tailored to specific FL tasks. CADRE generates comprehensive DR reports based on the user-defined metrics, rules, and remedies to ensure datasets are prepared for FL while preserving privacy. We demonstrate a practical application of CADRE by integrating it into an existing PPFL framework. We conducted experiments across six datasets and addressed seven different DR issues. The results illustrate the versatility and effectiveness of CADRE in ensuring DR across various dimensions, including data quality, privacy, and fairness. This approach enhances the performance and reliability of FL models as well as utilizes valuable resources.
\end{abstract}

\vspace{-10pt}
\section{Introduction}
\label{sec:introduction}
Federated Learning (FL) \cite{mcmahan2017communication,li2020federated} 
allows multiple decentralized participants to train a model collaboratively without sharing their raw data. Rather than centralizing data, FL allows each participant to locally train a model on their data and transmit only the model updates to a central server. This method enhances privacy and security by keeping sensitive data locally. However, new challenges emerge when privacy-preserving techniques are applied in FL. A recent study 
on Privacy-Preserving Federated Learning (PPFL) led by NIST \cite{nist_ppfl_data_pipeline_2024} 
highlights significant challenges, primarily due to the lack of access to training data.
Data cleaning and feature selection are complicated because data scientists cannot view data across different clients. This may lead to inconsistencies and deployment failures. Many studies \cite{nilsson2024impact,10710311,zhao2024enhancing} have demonstrated that low-quality data directly impacts the model by lowering the performance and robustness. Additionally, PPFL's privacy protections make it difficult to detect poor-quality or maliciously crafted data, which may lead to degrading the final model's quality. 
While recent research is beginning to address these issues with techniques like secure input validation and adaptations of data poisoning defenses \cite{chowdhury2022eiffelensuringintegrityfederated,cao2022fltrustbyzantinerobustfederatedlearning}, these solutions are not yet widely implemented in practical PPFL libraries.  


In our efforts to address these challenges, we introduce Data Readiness for AI (DRAI) into the PPFL domain. Our recent survey~\cite{10.1145/3722214} presented a comprehensive taxonomy for assessing DRAI, focusing on data quality, organization, fairness, understandability, governance, and value. We developed AIDRIN (AI Data Readiness Inspector) \cite{10.1145/3676288.3676296} framework to evaluate the DRAI of datasets across these dimensions. However, AIDRIN was initially designed for centralized AI training, where data is uploaded to a standalone platform for evaluation. In contrast, PPFL requires decentralized data readiness (DR) assessment, including methods that preserve privacy and security.

A framework for supporting user-defined metrics, rules, and remedies to allow data stewards and FL administrators to define custom metrics and evaluation criteria while preserving privacy is needed.
However, such a framework targeting either PPFL or centralized data readiness assessment is still unavailable. For example, in healthcare, PPFL can be used to develop a model for diagnosing a specific disease using MRI scans from multiple hospitals \cite{hoang2025enabling}. However, challenges such as data heterogeneity, quality, and privacy concerns arise because hospitals often use different MRI machines that leads to variations in image quality, resolution, and file format due to differences in hardware, software, and imaging protocols. In addition, some datasets may contain noisy or incomplete images, caused not only by machine differences but also by factors such as scanning artifacts, acquisition errors, or data corruption. To address these challenges, data owners should define custom DR standards, metrics, rules, and remedies tailored to their FL tasks. Data owners can establish standards for what constitutes an ``AI-ready'' MRI scan, such as data format and resolution requirements, and implement metrics to evaluate quality. Rules can be set to automatically flag images that do not meet these standards, and remedies, such as pre-processing techniques, can be applied to improve quality. 
It is required that each hospital ensure independently that its data meets the standards before participating in the FL process. 

To meet these challenging requirements in preparing and ensuring DR in PPFL, we propose a novel framework, called CADRE (\underline{C}ustomizable \underline{A}ssurance of \underline{D}ata \underline{RE}adiness). This framework allows FL ``administrators'' to define custom data readiness (DR) standards, including metrics, rules, and remedies tailored to specific FL tasks. Here, \textit{administrators} refers to the individuals or stakeholders responsible for a given FL task who collaborate to establish these data readiness standards. 
CADRE allows clients to locally execute these custom functions to ensure their data meets the necessary standards at run time without compromising privacy. Clients can verify compliance with these rules and apply remedies to their data as necessary. The results of these metric evaluations are compiled into a DR report for administrators' inspection. The report 
includes evaluations based on the custom readiness standards, along with standard metrics and visualizations of client data statistics. This framework brings a human-in-the-loop approach to FL by involving administrators in the definition, validation, and refinement of DR standards. CADRE can include predefined techniques and rules to showcase its capabilities, but its primary functionality lies in its customizability. The framework enables administrators to define a wide range of DRAI evaluation metrics, rules, and remedies tailored to specific FL tasks. This flexibility makes CADRE adaptable and practical for diverse PPFL scenarios, which enhances its usefulness across applications.

DR evaluation ensures that only clients with qualified data participate in the FL system. 
The CADRE framework is designed to be generalizable, applicable to any FL task, and adaptable to various domains. To demonstrate its practical application, we have developed an extensible module for the APPFL (Advanced Privacy-Preserving Federated Learning) framework \cite{li2024advances,ryu2022appfl}, an open-source software framework that enables researchers and developers to implement, test, and validate various PPFL techniques. With this integration, we showcase 
usage of CADRE in existing PPFL workflows. The main contributions of this study are:  


\begin{itemize}
    \item We propose a novel framework that enables FL administrators within a PPFL system to define custom metrics, rules, and remedies. CADRE addresses the execution of these custom standards by automating the process and ensuring that clients can locally apply these actions to meet required data standards while preserving privacy.
    \item We generate comprehensive DR reports in CADRE that evaluate the metrics defined by FL administrators. This ensures privacy preservation by only including aggregated metric evaluations without exposing any raw data. Administrators can review these reports to assess whether clients have met the expected standards and gain insights into the data's characteristics.
    \item We integrate CADRE into APPFL, demonstrating compatibility with existing PPFL workflows.

\end{itemize}

We evaluated CADRE using six datasets with a variety of data modalities (e.g., 2D images, tabular data, 3D volumetric data) and downstream tasks (such as classification, segmentation, and survival analysis). 
In some cases, we polluted the datasets to add noise, class imbalance, duplicate records, high memory consumption, bias, outliers, and insufficient anonymity. CADRE allows administrators within a PPFL system to define custom metrics, rules, and remedies, showing that the issues caused by our pollution were effectively addressed. We also demonstrate CADRE’s impact further using an example where resolving DR challenges leads to improvements in model performance. 
In the remainder of the paper, we describe related work (\S\ref{sec:related}), CADRE design (\S\ref{sec:design}), its integration into APPFL (\S\ref{sec:ppfl_integration}), and its evaluation (\S\ref{sec:eval}). 

\section{Related Work}
\label{sec:related}
\vspace{-3pt}
A few frameworks evaluate data 
with a focus on aspects such as data quality, governance, and infrastructure. Existing frameworks \cite{shrivastava2020dqlearn,gupta2021data,afzal2020data,schelter2018automating} primarily assess data availability, volume, quality, governance, and ethics. A wide range of data cleansing tools \cite{schelter2018automating,10.14778/3137628.3137631,ibm_dq_sla_compliance} are available today, each offering unique features to ensure the accuracy, reliability, and trustworthiness of data. However, most users prefer manual cleaning of the data and decide on AI readiness themselves or skip these tools entirely.

Despite their strengths, these frameworks exhibit critical gaps when applied to modern, distributed AI environments. They lack integration with FL architectures.
Existing frameworks generally assume a centralized data environment.
They also fall short in addressing compliance challenges related to cross-border data flows, which are common in FL scenarios. 

Ensuring the integrity of model updates is critical in FL, as malicious clients can degrade the quality of the global model. FLTrust \cite{cao2022fltrustbyzantinerobustfederatedlearning} addresses this by establishing a root of trust using a clean dataset to assign trust scores to client updates. However, its reliance on a single trusted dataset introduces a vulnerability if that dataset is compromised. EIFFeL \cite{chowdhury2022eiffelensuringintegrityfederated} enhances integrity while preserving privacy through secure aggregation and verification of client updates. It effectively filters out malicious contributions. However, it does not address the challenge of data heterogeneity, which can affect convergence and overall model performance.

The performance of FL models is often affected due to heterogeneous and noisy data distributions. In FL, where data is distributed between multiple clients, label noise refers to incorrect or inconsistent labels in the training data, which can significantly reduce model performance. To address this issue, FedELC \cite{Jiang_2024} proposes to identify clients with noisy labels and apply label correction strategies to refine the labels. However, it ignores other critical aspects of DR. FedDQA \cite{Zhang2024FedDQA} introduces a metric to evaluate client data quality, allowing the selection of higher-quality clients for training. Although effective in minimizing the influence of noisy data, this approach risks introducing selection bias and does not actively improve the underlying data. In the domain of PPFL, methods such as lazy influence approximation \cite{rokvic2024liaprivacypreservingdataquality} and FedDQC \cite{du2025dataqualitycontrolfederated} offer quality assessments that preserve privacy using influence scores and relevance alignment, respectively. Although these approaches maintain confidentiality, they have computational overhead and suffer from reduced data resolution under strict privacy constraints.

Another key limitation of existing FL frameworks is their lack of flexibility in supporting custom DR metrics and remediation workflows. Most rely on static, predefined evaluation criteria, making it difficult to accommodate domain-specific requirements. Remediation processes are often rigid and lack support for custom operations such as federated anonymization or edge-device preprocessing. Even unified data platforms rarely allow integration of custom rules or remedies. To address these gaps, our proposed framework enables FL administrators to define customized metrics, rules, and remedies aligned with the needs of specific FL systems. This flexibility helps manage data heterogeneity by enforcing consistent standards across clients, all while preserving privacy. The framework integrates seamlessly with existing PPFL workflows and supports DR evaluation before initiating resource-intensive training. Moreover, it aligns with the vision of Industry 5.0 \cite{eu_industry50}, emphasizing human-centric, privacy-aware, and adaptable AI systems that empower administrators to take control of DR.

\vspace{-4pt}
\section{Design Overview}
\label{sec:design}
\begin{figure}[htbp]
    \centering
    \includegraphics[width=\columnwidth]{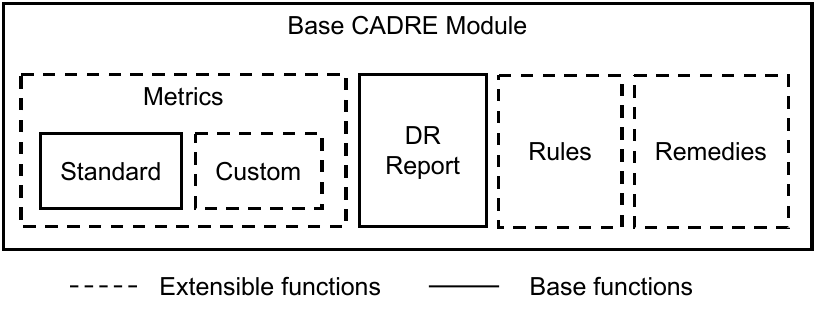}
    \caption{An overview of CADRE framework for FL tasks. Metrics include commonly known standard DR evaluation measurements. The extensible functions are used to define custom DR metrics, rules, and remedies. The DR report provides standard and custom metric evaluations with visualizations.}
    \label{fig:dr_agent_design}
\end{figure}

\vspace{-1pt}
The objective of CADRE is to allow administrators of FL systems to define and utilize both foundational and customizable actions.
To support this, CADRE provides the following main components: \emph{metrics}, \emph{DR reports}, \emph{rules}, and \emph{remedies}. 
In Figure \ref{fig:dr_agent_design}, we show an outline of CADRE with its components, including metrics that are standard and custom (i.e., administrator-defined). Rules and remedies are also administrator-defined functions. 

\vspace{-9pt}
\subsection{Metrics Component}
\vspace{-1pt}
We divided the metrics component of CADRE into two main parts: standard metrics and custom metrics. The standard metrics include a set of DR metrics defined in AIDRIN \cite{10.1145/3676288.3676296} including quality metrics such as evaluating sample sizes, data sparsity, and statistical measures like mean, median, and standard deviation of the client's data distribution. These metrics serve as a baseline for assessing DR of clients' data across any FL task. Additionally, the standard metrics component contains basic visualizations, such as bar charts and scatter plots, which are included in the DR reports to provide a visual representation of the client data characteristics. 

The custom metrics component is an extensible capability that allows FL administrators to provide custom metrics tailored to their unique FL task and evaluation needs. This flexibility ensures that administrators can assess client data according to the specific requirements of their projects. For example, if a task requires assessing the completeness or skewness of the data, administrators can define these metrics within CADRE. These standard and custom metric evaluations and visualizations allow administrators to quickly grasp the readiness of clients' data and identify potential issues that may lead to unexpected behavior in downstream FL tasks.

\subsection{Rules and Remedies Components}
Besides metrics, CADRE also includes sub-modules for defining rules and remedies. These sub-modules allow administrators to establish custom rules that their custom metric must meet to be considered ready for the next stages of the FL pipeline. The administrators can also define custom remedies to improve the readiness of the data to meet the specified rules. 
For instance, if administrators need to assess noise levels in the data, they can use metrics such as the standard deviation of the data distribution to quantify noise. A high standard deviation may indicate excessive variability and suggest the presence of noise. The administrators can then establish a rule where the standard deviation must not exceed a predefined threshold. If this threshold is surpassed, remedies could be implemented, such as filtering out extreme values or including only a subset of the affected client's data in the analysis.

\subsection{DR Reporting Module}
CADRE generates detailed DR reports by aggregating metric evaluations and visualizations produced by individual clients. It also includes principal component analysis (PCA) \cite{Pearson1901} graphs, to illustrate the combined data distribution and heterogeneity among clients. These insights are compiled into an easily readable HTML report, allowing administrators to assess whether clients meet specified standards while ensuring data privacy. This feature is essential to maintain transparency and accountability throughout the DR process.

\begin{figure}[htbp]
    \centering
    \includegraphics[width=.9\columnwidth]{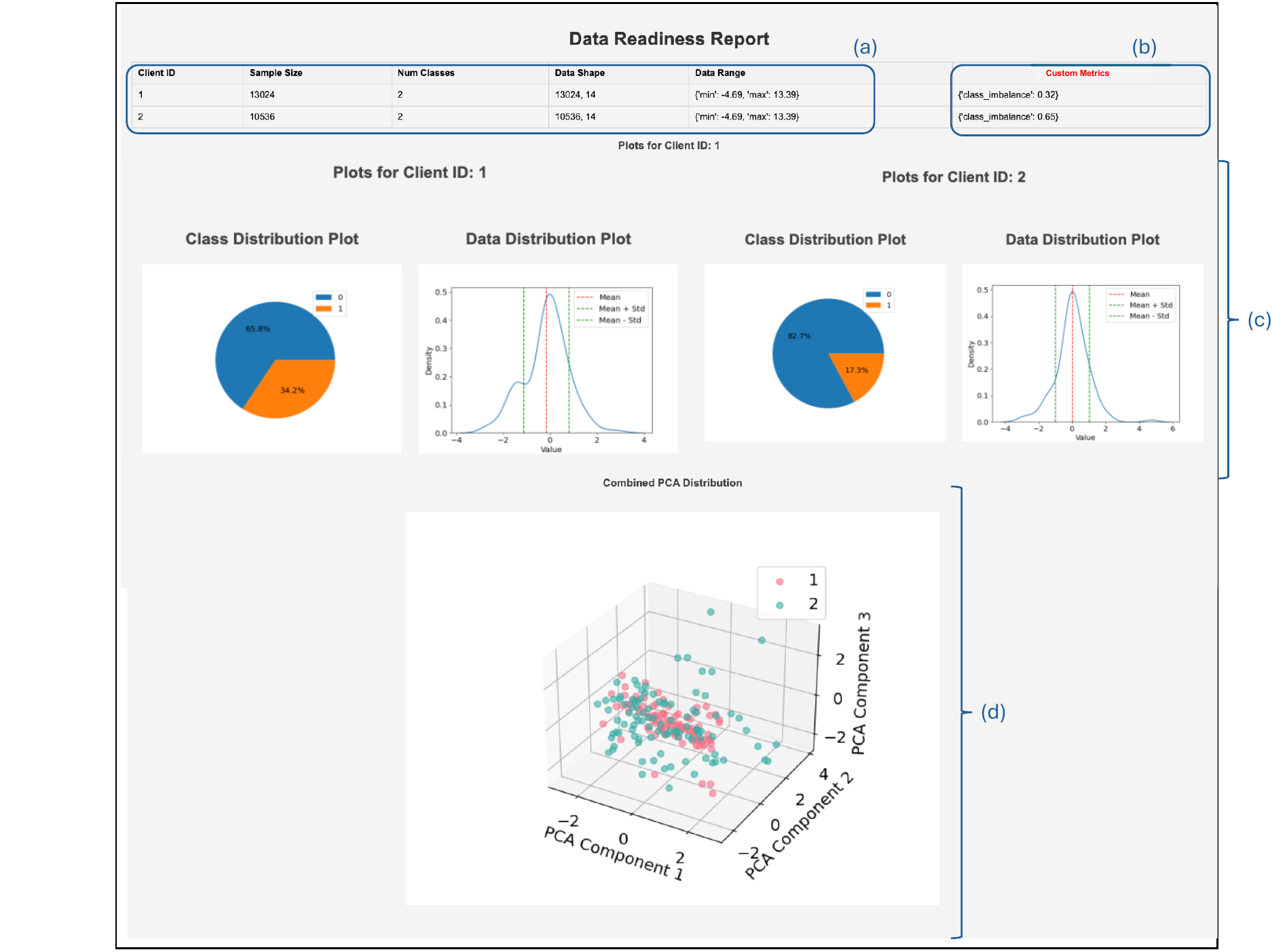}
    \caption{The figure illustrates an example DR report from an FL experiment featuring: (a) Standard metrics, (b) Custom metrics in CADRE for this specific FL task, (c) Individual client plots, and (d) Combined data plots.}
    \label{fig:dr_report_example}
\end{figure}

For instance, for a given FL task, a custom metric could involve measuring class imbalance within each client's dataset in the FL system. Identifying class imbalance is important because it can bias the learning process, especially in classification tasks where underrepresented classes may be poorly learned \cite{zhang2023surveyclassimbalancefederated}. In this scenario, a rule would be to flag any client datasets where the class distribution significantly deviates from a defined threshold of balance. If a client is flagged, the remedy might involve data augmentation or re-sampling techniques to mitigate the imbalance until the metric indicates an acceptable distribution. The resulting report will display the class distribution statistics for each client's dataset, making it easy to identify and address any flagged issues. In Figure \ref{fig:dr_report_example}, we present a DR report generated for this specific example. The report includes evaluations of both standard and custom metrics, visualizations for each of the two clients involved in the experiment, and combined plots. The visualizations include standard plots such as class distribution and data distribution charts, while the combined plot is a PCA visualization of a sample of the data from the clients. For this example, we used the Adult Income dataset \cite{Kohavi1996}, and CADRE is integrated into the APPFL framework. More details about this integration and the experiments can be found in sections \ref{sec:ppfl_integration} and \ref{sec:eval}.

\begin{figure}[htbp]
    \centering
    \includegraphics[width=\columnwidth]{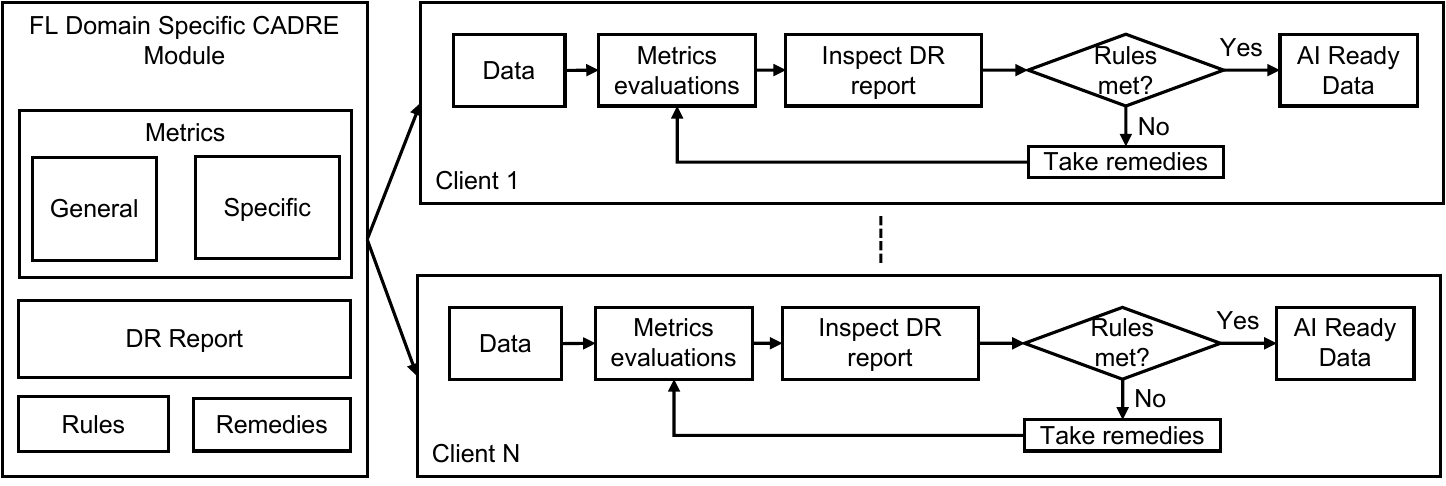}
    \caption{An iterative data evaluation and remediation process where clients are involved in an FL framework. It outlines how clients use the CADRE's functions to assess DR, apply custom rules, and implement remedies as needed.}
    \label{fig:dr_agent_usage}
\end{figure}

Clients participating in the FL framework use custom metrics within CADRE to locally evaluate their data and generate DR reports. If the client data meets the specified rules, the data will proceed to the subsequent stages of the FL pipeline. Conversely, if the data does not meet the rules, remedies defined by the administrators within CADRE will be applied to improve the DR. This process will iterate until the data complies with the established rules. This will ensure DR for the next stages of the FL pipeline. Figure~\ref{fig:dr_agent_usage} provides a visual representation of this iterative approach by illustrating how clients use CADRE's functions to assess DR, apply custom rules, and implement remedies while preserving privacy.

By integrating metrics, rules, and remedies components into DR frameworks, CADRE provides a comprehensive and flexible framework for ensuring DR in FL systems. This framework allows administrators to tailor the DR process to the specific needs of their projects while maintaining high standards of DR and privacy.


\section{Integration Into Existing PPFL Frameworks}
\label{sec:ppfl_integration}

In this study, we utilize the APPFL framework to demonstrate the practical application of CADRE. APPFL is an open-source framework designed to enhance privacy and security in FL systems. It allows researchers to implement, test, and deploy FL experiments across distributed clients while ensuring data privacy. We chose APPFL as the testbed for CADRE due to its modular and extensible architecture, which aligns well with CADRE’s design principles and integration goals. Its built-in support for differential privacy, asynchronous and synchronous training algorithms, and flexible customization of core FL components makes it a suitable and practical platform for evaluating CADRE’s capabilities in real-world PPFL scenarios.

APPFL consists of six key components: an aggregator, scheduler, trainer, privacy module, communicator, and compressor. These components work together to tackle challenges such as computational disparities and security concerns in distributed machine learning, while also enabling enhanced privacy protection, supporting flexible model training on decentralized data, simulating various FL algorithms, implementing lossy compression for efficient data transfer, and providing a highly extensible framework for customizing aggregation algorithms, server scheduling strategies, and client local trainers. The framework supports various popular synchronous and asynchronous FL algorithms such as FedAvg \cite{mcmahan2017communication}, FedAvgM \cite{fedavgm}, FedBuff \cite{nguyen2022federated}, and FedCompass \cite{li2023fedcompass}, and incorporates differential privacy techniques \cite{dwork2006differential}.

\begin{figure}[htpb]
    \centering
    \includegraphics[width=\columnwidth]{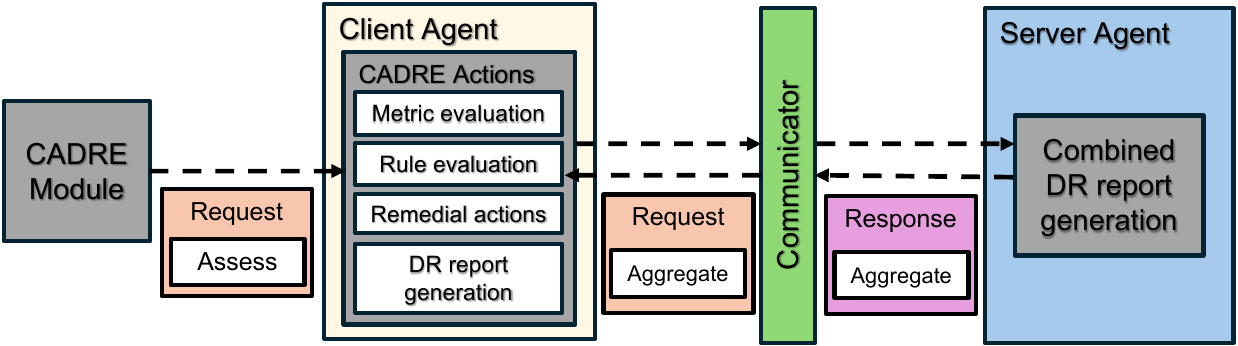}
    \caption{An integration of CADRE in the APPFL framework.}
    \label{fig:appfl_integration}
\end{figure}

CADRE will be integrated into the APPFL framework as an extensible module. Administrators can use the extensible nature of CADRE to define the metrics, rules, and remedies for a specific FL task. This allows clients to use its functions locally. This integration enables clients to evaluate data using custom metrics and apply custom remedies if the rules are not satisfied. After evaluating the data, the client agent will compile a DR report of the evaluations. These evaluations are then aggregated by the communicator within APPFL to combine the results from all clients for review. This integration demonstrates CADRE's ease of use and versatility within existing PPFL frameworks. In Figure~\ref{fig:appfl_integration}, we provide an overview of its implementation within the APPFL framework.

Configuring CADRE for specific FL tasks is a straightforward process that allows administrators to tailor its extensible functionality to meet the unique requirements of each task. The process begins with the utilization of the base CADRE module. The base CADRE module serves as a foundational template with extensible functions. By using this template, administrators can create a specialized CADRE module that incorporates the necessary evaluation metrics, rules, and remedies specific to their task.

\begin{figure}[h]
    \centering
    \includegraphics[width=0.6\columnwidth]{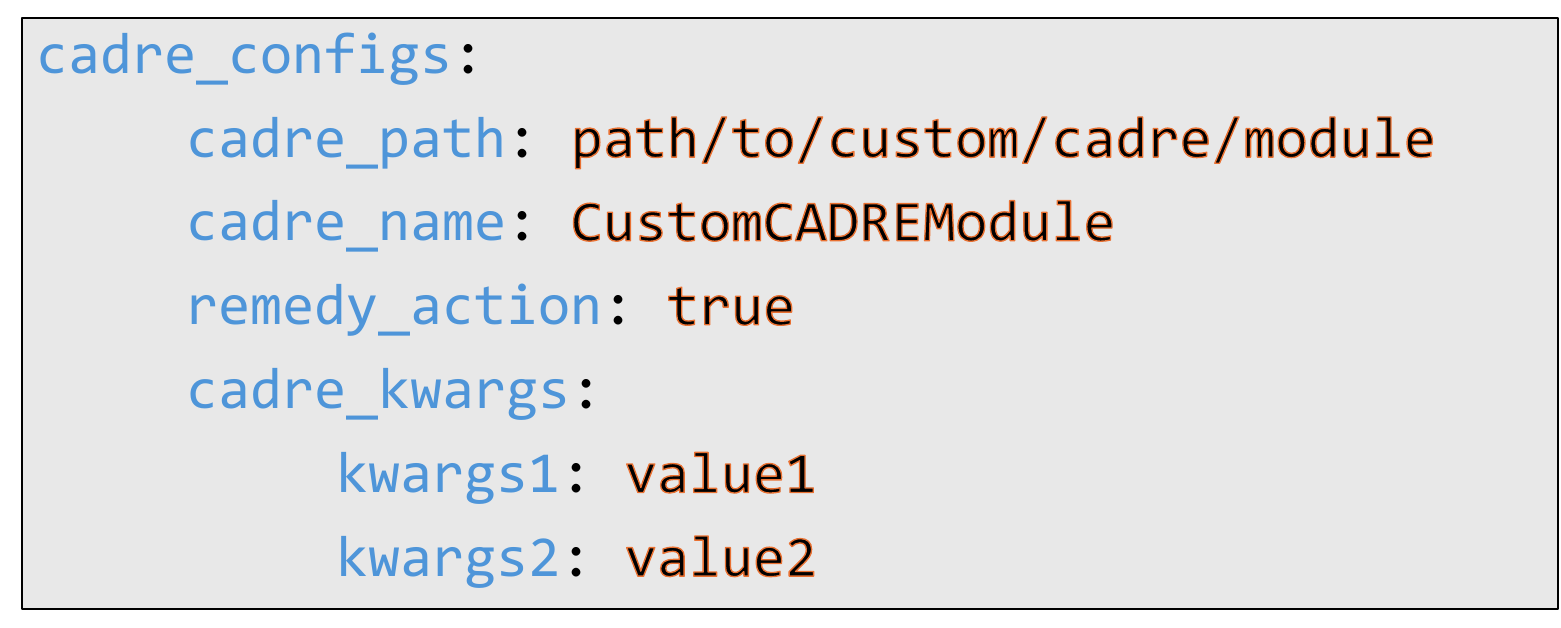}
    \caption{YAML configuration for customizing a CADRE module in FL tasks, allowing administrators to define evaluation metrics, rules, and remedies specific to their needs.}
    \label{fig:yaml_config}
\end{figure}

Once the custom CADRE module is configured, it is seamlessly integrated into the APPFL framework by uploading it. The framework is designed to accommodate such modular additions, making the integration process smooth and efficient. To activate the newly created CADRE module, administrators simply update the configuration file within the APPFL framework. This involves specifying the path to the custom CADRE module file that will allow the framework to recognize and utilize it appropriately. Additionally, administrators can pass other relevant arguments specific to the CADRE module by defining them in the configuration file. For instance, a CADRE module may require additional inputs, such as feature indices and other identifiers, for various DR-related tasks. Figure \ref{fig:yaml_config} illustrates an example of this configuration, showcasing the YAML-based setup used to define a custom CADRE module.

With the integration of CADRE into the APPFL framework, administrators gain significant advantages that aid in making informed decisions before entering the costly training phase. As data flows through the system, CADRE automatically executes defined actions, ensuring that DR issues are addressed promptly and consistently. This automation provides administrators with timely interventions, allowing them to focus on strategic decisions rather than manual data remediation tasks.

Additionally, the DR reports offer transparency and accountability. These reports provide administrators with a clear overview of the DR actions taken and allow effective assessment of DR compliance. By reviewing the detailed evaluations without exposing any raw data, while maintaining privacy and security,  administrators can ensure that only clean and compliant data is used. Overall, this streamlined approach highlights how easily CADRE can be adapted for different FL tasks and data modalities. This concept will enhance the flexibility and effectiveness of PPFL. The documentation and code for this integration are available as part of APPFL \cite{appflIntegration}.

Integration of CADRE into APPFL leads to improved model performance, as AI-ready data reduces the risk of errors and noise affecting the training process. CADRE also supports scalability by allowing the system to efficiently handle large datasets. This allows administrators to make better-informed decisions, optimizing resource allocation and minimizing risks before committing to the next phases in FL.

\vspace{-2pt}
\section{Evaluations}
\vspace{-2pt}
\label{sec:eval}
To demonstrate the effectiveness of CADRE in evaluating data quality, privacy, and fairness, we use multiple datasets, experimental setups, and custom CADRE modules.
Since most of the publicly available datasets that have been used to develop FL models are relatively clean and preprocessed,
we used various data pollution techniques to evaluate with CADRE.
We will illustrate how our custom DR standards are achieved by utilizing the tailored metrics, rules, and remedies within the custom CADRE modules. We will present an example illustrating the performance improvement of the final FL model on the downstream task when using CADRE, compared to without using it.

\begin{table*}
\caption{Overview of custom CADRE modules used in experiments.}
\label{tab:agent_desc}
\centering
\resizebox{\textwidth}{!}{
\begin{tabular}{|c|p{3cm}|p{4cm}|p{5cm}|p{5cm}|}
\hline
\textbf{CADRE Module ID} & \textbf{Category} & \textbf{Metric} & \textbf{Rule} & \textbf{Remedy} \\
\hline
1 & Noise Management & Mean magnitude of the data (image intensities or feature values) & Applied remedy when the data distribution mean exceeded a threshold (e.g., $> 0.37$ for MNIST). & Data points with noisy indices were removed. \\
\hline
2 & Class Imbalance Handling & Class imbalance degree \cite{xiao2021experimental} & Applied when imbalance degree $> 0$. & SMOTE \cite{Chawla2002SMOTE} was used to oversample the minority class. \\
\hline
3 & Duplicate Management & Proportion of duplicates & Applied when duplicates proportion $> 0$. & Duplicates were identified and removed. \\
\hline
4 & Memory Optimization & Memory usage in megabytes (MB) to store the client's data & Applied when memory usage was excessively high. & Data types were optimized or duplicates removed depending on the dataset’s pollution method. \\
\hline
5 & Bias Handling & Statistical parity difference \cite{Dwork2012Fairness} for Adult Income dataset and representative rate difference for TCGA-BRCA dataset & Applied when metric value $> 0$. & Stratified resampling \cite{liberty2016stratified} to balance sensitive groups and labels in the Adult Income dataset, while SMOTE to oversample the minority group in the TCGA-BRCA dataset. \\
\hline
6 & Outlier Management & Proportion of outliers using Inter-quartile range (IQR) method \cite{Tukey1977} & Applied when outliers proportion $> 0$. & Outliers were clipped at IQR bounds. \\
\hline
7 & K-anonymity Handling & K-anonymity level \cite{Samarati1998Protecting} & Applied when anonymity level $\leq 1$. & Data records with low anonymity levels were suppressed to ensure the desired level of anonymity. \\
\hline
\end{tabular}
}
\end{table*}

\vspace{-4pt}
\subsection{Datasets and Experimental Setup}
\vspace{-2pt}

In this study, we used six datasets spanning both standard benchmarks and those from real-world medical research. The benchmark datasets include MNIST \cite{lecun2010mnist}, a collection of handwritten digit images widely used for image classification; CIFAR-10 \cite{Krizhevsky09learningmultiple}, which comprises color images across ten classes for object recognition tasks; and Adult Income \cite{Kohavi1996}, a tabular dataset from the UCI repository used to predict whether an individual's income exceeds \$50K based on census data.

In addition to these, we used three datasets derived from real-world medical research. TCGA-BRCA from the Flamby collection \cite{he2022flamby} contains clinical data from breast cancer patients and is used for survival analysis. The IXI Tiny dataset, also from Flamby, consists of 3D brain MRI scans and serves as a benchmark for medical image segmentation tasks. Both of these datasets are naturally partitioned among clients, such as different hospitals or research centers, and are widely used in FL research. Finally, the AI-READI (Artificial Intelligence Ready and Equitable Atlas for Diabetes Insights) dataset \cite{AI-READI_Consortium_2024} is a new comprehensive and ethically sourced collection designed to advance AI research in Type 2 Diabetes Mellitus (DM2), consisting over 15 data modalities, such as vitals, retinal imaging, electrocardiograms, and other health-related measurements, all aimed at exploring salutogenic pathways to health. For our research, we utilized color fundus photography (CFP) images from the AI-READI collection to classify the severity of diabetes by analyzing the retinal health using the CFP images. To simulate real-world heterogeneity, we divided the dataset among four clients based on the imaging devices used: iCare Eidon, Optomed Aurora, Topcon Maestro2, and Topcon Triton. 
By considering these datasets from various modalities and with different downstream tasks, we demonstrate the versatility of our proposed framework, which is not constrained by data modality or task.

To facilitate the evaluation of class imbalance, we transformed MNIST, CIFAR-10, and the AI-READI data into binary classification tasks. In MNIST, digits 0–4 were grouped into one class, while digits 5–9 formed another. In CIFAR-10, images with class indices 0–4 were assigned to one class, while images with class indices 5–9 were categorized as the other.  For the AI-READI dataset, we categorized the classes as follows: the ``pre-diabetes (lifestyle controlled)'' and ``oral medication and/or non-insulin injectable medication controlled'' classes were combined into one group, while the ``healthy'' and ``insulin-dependent'' classes formed the other group. This transformation simplifies the evaluation process and improves understandability. The Adult Income dataset is inherently a binary classification task, so no further modifications were necessary.

As discussed in section \ref{sec:ppfl_integration}, we employed APPFL to integrate CADRE and conduct the experiments. We consistently used FedAvg \cite{mcmahan2017communication} as the primary FL algorithm across all experiments. Since MNIST, CIFAR-10, and Adult Income are not inherently FL datasets, we applied non-independent and identically distributed (non-IID) partitioning to ensure data heterogeneity. 
For these three datasets, we partitioned the data into 10 clients per experiment and ran the experiments for 10 global epochs. On the other hand, TCGA-BRCA and IXI Tiny datasets are genuine FL datasets, already partitioned into 6 and 3 clients, respectively. As previously mentioned, the AI-READI dataset was partitioned based on the imaging device used, resulting in four clients corresponding to the four devices. CADRE operates before the actual training phase, so FL training related configurations do not impact CADRE's execution. However, to ensure the completeness of our experiments and to validate integration in FL tasks, we reported these configurations. For the AI-READI dataset, we utilized a single node with 64GB RAM and one NVIDIA A40 GPU on the Delta supercomputer at NCSA \cite{gropp2023delta}. The rest of the experiments were conducted on an Apple M2 Max MacBook Pro with 32GB unified memory. 

\vspace{-3pt}
\subsection{Custom CADRE Modules}
\vspace{-3pt}
In this study, we used seven custom CADRE modules, each designed to address a specific DR issue. These modules incorporate tailored metrics, rules, and remedies to ensure that the client's data meets the expected standards. The selection of modules covers a broad spectrum of DR challenges, as identified in the \cite{10.1145/3722214} study, including data quality, fairness, privacy, and structure. Table \ref{tab:agent_desc} provides a detailed overview of these custom modules by outlining the metrics, rules, and remedies each module uses to evaluate and enhance the data's readiness for specific AI tasks.

\begin{table*}
\caption{Dataset-specific data pollution methods applied for each CADRE module.}
\label{tab:dataset_pollution}
\centering
\resizebox{\textwidth}{!}{
\begin{tabular}{|c|p{3.5cm}|p{3.5cm}|p{3.5cm}|p{3.5cm}|p{3.5cm}|p{3.5cm}|}
\hline
\textbf{CADRE Module ID} & \textbf{MNIST} & \textbf{CIFAR-10} & \textbf{Adult Income} & \textbf{Flamby TCGA-BRCA} & \textbf{Flamby IXI Tiny} & \textbf{AI-READI} \\ 
\hline
1 & Added Gaussian noise (std. dev. = 2) to 90\% of the data & Added Gaussian noise (std. dev. = 2) to 90\% of the data & Added Gaussian noise (std. dev. = 2) to 90\% of the data & Added Gaussian noise (std. dev. = 2) to 90\% of the data & Added Gaussian noise (std. dev. = 2) to 90\% of the data & Added Gaussian noise (std. dev. = 2) to 90\% of the data \\
\hline
2 & Imbalanced class distribution due to non-IID partitioning & Imbalanced class distribution due to non-IID partitioning & Imbalanced class distribution due to non-IID partitioning & Not applicable (survival analysis task) & Not applicable (segmentation task) & Device-based partitioning inherently resulted in an imbalanced class distribution \\
\hline
3 & 20\% of data was randomly duplicated & 20\% of data was randomly duplicated & 20\% of data was randomly duplicated & 20\% of data was randomly duplicated & 20\% of data was randomly duplicated & 20\% of data was randomly duplicated \\
\hline
4 & Converted feature values to higher precision (float32 to float64) & Converted feature values to higher precision (float32 to float64) & Converted feature values to higher precision (float32 to float64) & Duplicates added to increase memory usage & Duplicates added to increase memory usage & Duplicates added to increase memory usage \\
\hline
5 & Not applicable (image data has no sensitive features) & Not applicable (image data has no sensitive features) & Statistical parity differences were inherent & Representative rate differences were inherent & Not applicable (image data has no sensitive features) & Not applicable (image data has no sensitive features) \\
\hline
6 & Added random gaussian noise (std. dev. = 2) to the data to simulate outliers & Added random gaussian noise (std. dev. = 2) to the data to simulate outliers & Added random gaussian noise (std. dev. = 2) to the data to simulate outliers & Features inherently contained outliers & Added random gaussian noise (std. dev. = 2) to the data to simulate outliers & Added random gaussian noise (std. dev. = 2) to the data to simulate outliers \\
\hline
7 & Not applicable (no quasi-identifiers in image data) & Not applicable (no quasi-identifiers in image data) & Quasi-identifiers already contained low levels of anonymity & Quasi-identifiers already contained low levels of anonymity & Not applicable (no quasi-identifiers in image data) & Not applicable (no quasi-identifiers in image data) \\
\hline
\end{tabular}
}
\end{table*}

As seen in Table~\ref{tab:agent_desc}, for module 5, we measured statistical parity difference in the Adult Income dataset and representation rates in the TCGA-BRCA dataset. Statistical parity involves assessing class labels and sensitive groups, making it suitable for the Adult Income dataset, which deals with classification tasks. However, for the TCGA-BRCA dataset, which is used for survival analysis, measuring statistical parity is not feasible. Instead, we evaluate the representation rates of sensitive attributes and balance them as a remedy. For the Adult Income dataset, ``gender'' was selected as the sensitive feature for analysis by the module. This feature contains two categories: ``male'' and ``female.'' In contrast, for the TCGA-BRCA dataset, ``race\_white'' was identified as the sensitive feature, represented as a binary attribute where ``1'' indicates that the race is white, and ``0'' signifies otherwise.

\begin{figure*}[htbp]
    \centering
    \includegraphics[width=\textwidth]{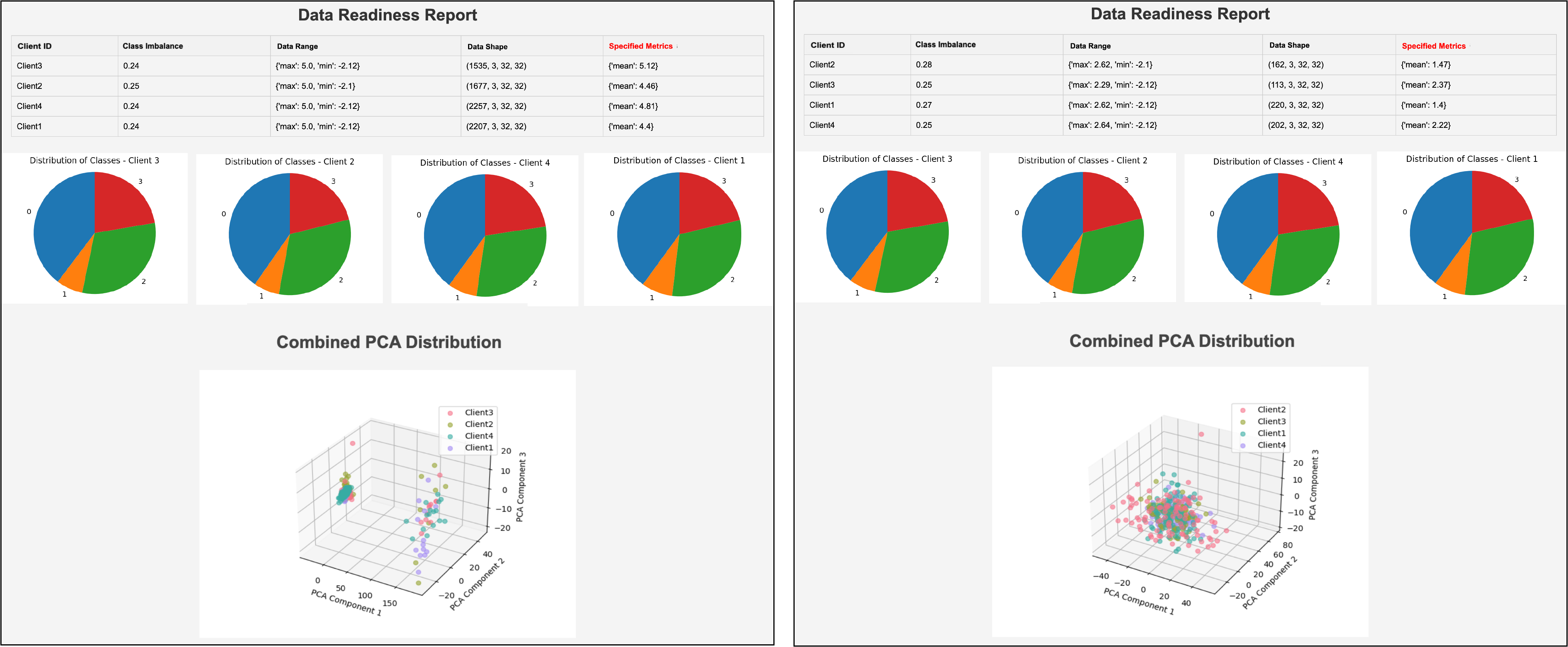}  
    \captionsetup{belowskip=-17pt} 
    \caption{Example DR reports generated before (left) and after (right) applying CADRE module 1 show an improvement in the average mean after removing noisy data. Results are shown in the table's rightmost column. The combined PCA plot at the bottom right confirms that noise-related anomalies in the data distribution have been resolved.}
    \label{fig:aidrin_report_before_after} 
\end{figure*}

Module 7 uses k-anonymity level as a metric. A remedy is applied when the anonymity level is less than or equal to 1 by ensuring that each entity remains identical from at least $k-1$ others based on quasi-identifiers \cite{Samarati1998Protecting}. Quasi-identifiers are attributes that are not unique identifiers on their own but can be combined to identify individuals. For the Adult Income dataset, quasi-identifiers were ``workclass,'' ``race,'' and ``gender.'' We selected these as the quasi-identifiers because they are commonly available in public records and, when combined, could increase re-identification risk. Similarly, for the TCGA-BRCA dataset, the quasi-identifiers included demographic and self-reported characteristics such as ``age\_at\_index,'' ``ethnicity\_not hispanic or latino,'' ``ethnicity\_not reported,'' ``race\_asian,'' ``race\_black or african american,'' ``race\_not reported,'' and ``race\_white.'' These attributes were chosen due to their potential to link individuals across datasets and may pose privacy concerns if identified.

\vspace{-2pt}
\subsection{Data Pollution}
\vspace{-2pt}
To fully demonstrate the remedies provided by our custom CADRE modules, it was essential to ensure that the datasets used in our study exhibited the relevant issues. Some datasets naturally contained issues such as class imbalance, which was present in all classification tasks due to non-IID partitioning. Other issues were intentionally introduced through data pollution techniques. Table \ref{tab:dataset_pollution} provides detailed information on the pollution methods applied to each dataset. By polluting data, it enables the activation of rule and remedy actions in the custom CADRE modules in every experiment.

Figure \ref{fig:aidrin_report_before_after} presents two DR report samples from an experiment conducted before and after meeting a CADRE module's standards. These reports illustrate how easily data-related issues can be identified and addressed, ensuring that standards defined by the custom CADRE modules are met. For this sample, we used the AI-READI dataset's before-and-after DR reports from the experiment conducted for CADRE module 1.

\vspace{-2pt}
\subsection{Results}
\vspace{-3pt}

\begin{figure*}[htbp]
    \centering
    \includegraphics[width=\textwidth]{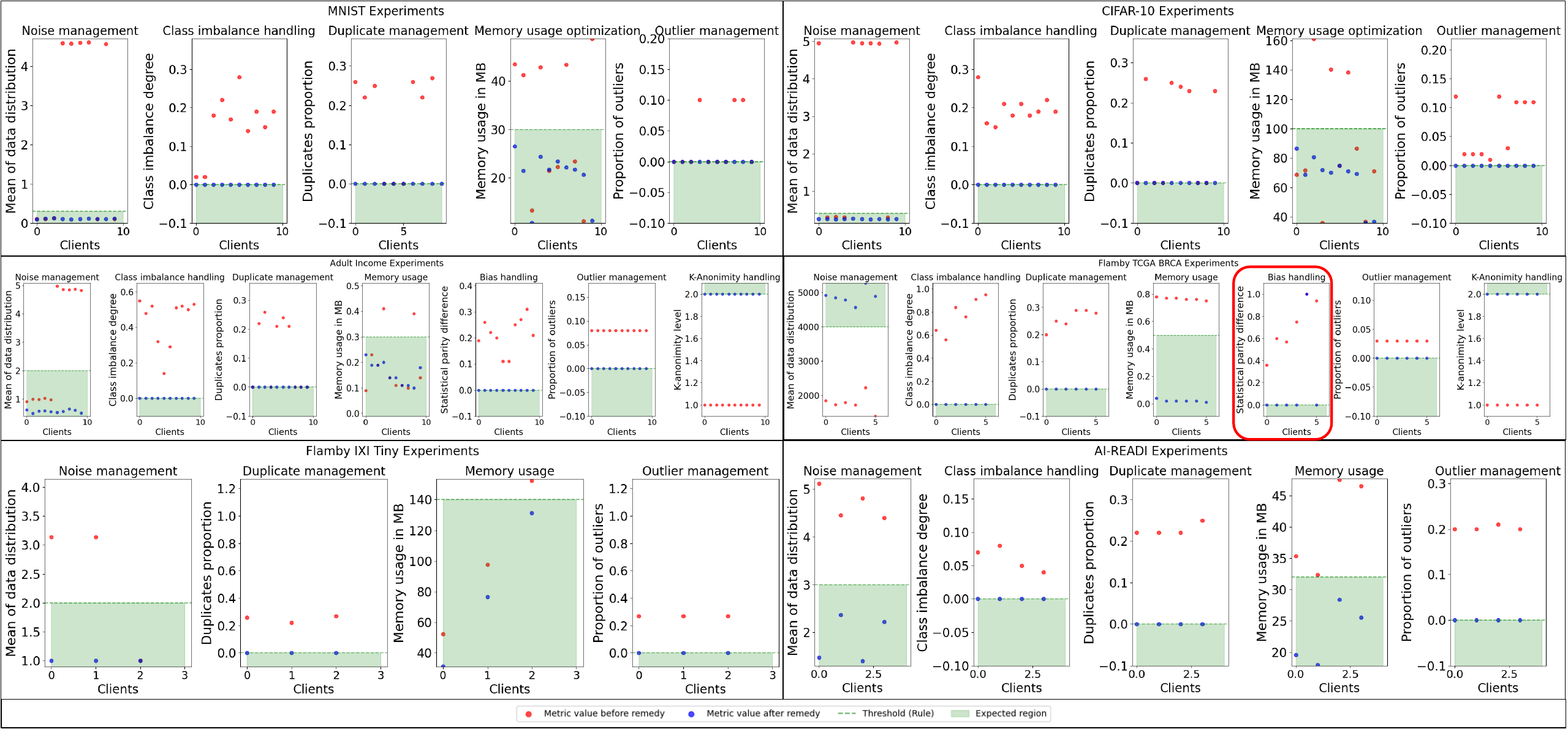}  
    \captionsetup{belowskip=-17pt} 
    \caption{Evaluation of custom metrics for each CADRE module, before and after remedy application. Threshold lines indicate predefined rule criteria. The red box highlights one case where a client's post-remedy metric remains above the threshold.}
    \label{fig:results}  
\end{figure*}
After conducting experiments across all datasets and custom CADRE modules, as detailed in Tables \ref{tab:agent_desc} and \ref{tab:dataset_pollution}, we observed that nearly all client data met the required standards defined by each custom CADRE module. The process generated DR reports that reflected these metric evaluations, along with standard metrics and visualizations, as depicted in the Figure \ref{fig:aidrin_report_before_after}. Figure \ref{fig:results} illustrates the metric values before and after applying the remedies of custom CADRE modules, with threshold values indicating the rules set for each experiment. As shown in the figure, almost all post-remedy data points fall within the expected range. However, there is only one exception, observed in the figure that is boxed in red, where one client's post-remedy metric value remains above the threshold. The DR report's representative rates plot of the sensitive feature helped identify that this particular client contained only one ethnic group, preventing the remedy action from balancing the feature due to the absence of a second group. This example highlights the importance of DR reports in understanding the DR levels of clients before proceeding to the training phase.


Although this work focuses on the pre-training phase of the FL pipeline, it offers important insights into the quality and readiness of data before initiating costly training procedures. By evaluating and improving DR early on, administrators can make informed decisions about whether to proceed with training. This will ultimately help conserving computational and organizational resources.

\begin{figure}[htbp]
    \centering
    \includegraphics[width=\columnwidth]{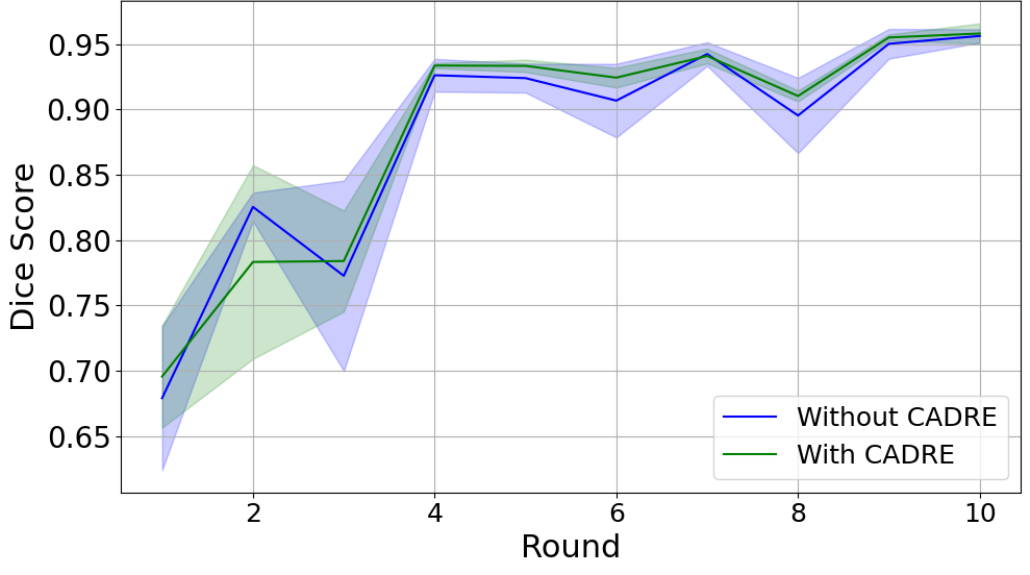}
    \caption{The figure presents a comparative analysis of FL performance on the Flamby IXI Tiny dataset by evaluating the impact of CADRE's noise handling module through Dice scores over ten training rounds.}
    \label{fig:performance_results}
\end{figure}

To demonstrate the downstream impact of CADRE, we conducted an experiment using the IXI Tiny dataset from the Flamby benchmark suite. This dataset consists of 3D brain MRI scans and is commonly used for medical image segmentation tasks, where performance is typically measured using the Dice score \cite{flamby_ixi_dice}. Figure \ref{fig:performance_results} presents the average Dice scores between clients during 10 rounds of FL training. The blue curve shows performance before applying the CADRE noise management module, while the green curve reflects results after CADRE was used to remove noisy indices. The shaded regions represent the standard deviation between clients to capture the variability in performance.

The results indicate that applying CADRE leads to consistently improved and more stable Dice scores, particularly from round 4 onward. This suggests that CADRE’s early noise-handling improves DR and reduces inter-client variability. These factors are critical for achieving better generalization in FL models. These findings align with prior research showing that high-quality, noise-free data improves model robustness and accuracy \cite{nilsson2024impact,rokvic2024liaprivacypreservingdataquality}, and that addressing other DR dimensions, such as class imbalance, can mitigate bias and reduce model drift \cite{Xiao2022Experimental,risingwave_federated_learning_2023}.

However, other factors, such as achieving perfect fairness and optimal anonymity levels, may affect different aspects of model performance. A dataset with minimal statistical parity can improve model fairness \cite{Huang2022Fairness}, though it may compromise overall performance and accuracy. Similarly, as we increase the privacy budget of the data, model accuracy tend to decrease \cite{Fisichella2022Partially}. However, administrators might choose to prioritize data fairness, and privacy standards over model performance. Also, memory usage optimization is crucial for FL clients, as resource-constrained edge devices have limited computational and memory capacity \cite{Huang2024FedMef}. Efficient optimization helps maintain training efficiency while preventing performance degradation. Overall, these results demonstrate that our framework can be effectively integrated into PPFL systems to meet DR-related standards before training to conserve valuable resources and funds. Moreover, the informative DR reports simplify the process for administrators by providing a clear understanding of the data's condition for the FL task and setting expectations for the training phase.

\section{Conclusion and Future Work}
\label{sec:conclusion}
In this study, we introduced a novel framework to enhance DR in PPFL systems. The framework allows FL administrators to define CADRE modules tailored to address diverse DR challenges across various downstream tasks and data modalities. By specifying custom metrics, rules, and remedies, these modules allow clients to execute processes locally and to ensure that their data meets the necessary standards while preserving privacy. 
CADRE allows administrators to set realistic expectations for training, optimize resource utilization, and lay the groundwork for reliable and equitable FL results.

In our future work, we will expand CADRE's applicability to a broader range of usecases and explore automated methods to streamline the DR process. Additionally, we will investigate computationally intensive tasks and explore adding custom privacy-preserving modules to CADRE for user-controlled privacy protection. This will enhance the adaptability of CADRE to evolving privacy standards in PPFL frameworks.

\section*{Acknowledgment}
\label{sec:acknowledgments}
This work is supported in part by the U.S. Department of Energy, Office of Science, under contract numbers DE-AC02-06CH11357, DE-AC02-05CH11231, and subcontracts GR138942 and GR130493 at OSU. This research uses computing resources provided by the National Artificial Intelligence Research Resource (NAIRR) Pilot, supported by award NAIRR240008.

\bibliographystyle{IEEEtran}
\bibliography{references}

\begin{thebibliography}{10}
\providecommand{\url}[1]{#1}
\csname url@samestyle\endcsname
\providecommand{\newblock}{\relax}
\providecommand{\bibinfo}[2]{#2}
\providecommand{\BIBentrySTDinterwordspacing}{\spaceskip=0pt\relax}
\providecommand{\BIBentryALTinterwordstretchfactor}{4}
\providecommand{\BIBentryALTinterwordspacing}{\spaceskip=\fontdimen2\font plus
\BIBentryALTinterwordstretchfactor\fontdimen3\font minus \fontdimen4\font\relax}
\providecommand{\BIBforeignlanguage}[2]{{%
\expandafter\ifx\csname l@#1\endcsname\relax
\typeout{** WARNING: IEEEtran.bst: No hyphenation pattern has been}%
\typeout{** loaded for the language `#1'. Using the pattern for}%
\typeout{** the default language instead.}%
\else
\language=\csname l@#1\endcsname
\fi
#2}}
\providecommand{\BIBdecl}{\relax}
\BIBdecl

\bibitem{mcmahan2017communication}
B.~McMahan, E.~Moore, D.~Ramage, S.~Hampson, and B.~A. y~Arcas, ``Communication-efficient learning of deep networks from decentralized data,'' \emph{Artificial intelligence and statistics}, pp. 1273--1282, 2017.

\bibitem{li2020federated}
T.~Li, A.~K. Sahu, A.~Talwalkar, and V.~Smith, ``Federated learning: Challenges, methods, and future directions,'' \emph{IEEE Signal Processing Magazine}, vol.~37, no.~3, pp. 50--60, 2020.

\bibitem{nist_ppfl_data_pipeline_2024}
X.~Huang, Y.~Dong, and S.~Pentyala, ``Data pipeline challenges in privacy-preserving federated learning,'' \url{https://www.nist.gov/blogs/cybersecurity-insights/data-pipeline-challenges-privacy-preserving-federated-learning}, February 2024, nIST Cybersecurity Insights Blog Post. Part of a series on privacy-preserving federated learning in collaboration with the UK government’s Responsible Technology Adoption Unit (RTA).

\bibitem{nilsson2024impact}
G.~Nilsson, ``The impact of data quality on federated versus centralized learning,'' Master of Science in Engineering: AI and Machine Learning, Blekinge Institute of Technology, 371 79 Karlskrona, Sweden, July 2024.

\bibitem{10710311}
G.~Nilsson, M.~Boldt, and S.~Alawadi, ``The role of the data quality on model efficiency: An exploratory study on centralised and federated learning,'' in \emph{2024 9th International Conference on Fog and Mobile Edge Computing (FMEC)}, 2024, pp. 253--260.

\bibitem{zhao2024enhancing}
W.~Zhao, Y.~Du, N.~D. Lane, S.~Chen, and Y.~Wang, ``Enhancing data quality in federated fine-tuning of foundation models,'' \emph{arXiv preprint arXiv:2403.04529}, Mar 2024.

\bibitem{chowdhury2022eiffelensuringintegrityfederated}
\BIBentryALTinterwordspacing
A.~R. Chowdhury, C.~Guo, S.~Jha, and L.~van~der Maaten, ``Eiffel: Ensuring integrity for federated learning,'' 2022. [Online]. Available: \url{https://arxiv.org/abs/2112.12727}
\BIBentrySTDinterwordspacing

\bibitem{cao2022fltrustbyzantinerobustfederatedlearning}
\BIBentryALTinterwordspacing
X.~Cao, M.~Fang, J.~Liu, and N.~Z. Gong, ``Fltrust: Byzantine-robust federated learning via trust bootstrapping,'' 2022. [Online]. Available: \url{https://arxiv.org/abs/2012.13995}
\BIBentrySTDinterwordspacing

\bibitem{10.1145/3722214}
\BIBentryALTinterwordspacing
K.~Hiniduma, S.~Byna, and J.~L. Bez, ``Data readiness for ai: A 360-degree survey,'' \emph{ACM Comput. Surv.}, Mar. 2025, just Accepted. [Online]. Available: \url{https://doi.org/10.1145/3722214}
\BIBentrySTDinterwordspacing

\bibitem{10.1145/3676288.3676296}
\BIBentryALTinterwordspacing
K.~Hiniduma, S.~Byna, J.~L. Bez, and R.~Madduri, ``Ai data readiness inspector (aidrin) for quantitative assessment of data readiness for ai,'' in \emph{Proceedings of the 36th International Conference on Scientific and Statistical Database Management}, ser. SSDBM '24.\hskip 1em plus 0.5em minus 0.4em\relax New York, NY, USA: Association for Computing Machinery, 2024. [Online]. Available: \url{https://doi.org/10.1145/3676288.3676296}
\BIBentrySTDinterwordspacing

\bibitem{hoang2025enabling}
T.-H. Hoang, J.~Fuhrman, M.~Klarqvist, M.~Li, P.~Chaturvedi, Z.~Li, K.~Kim, M.~Ryu, R.~Chard, E.~A. Huerta \emph{et~al.}, ``Enabling end-to-end secure federated learning in biomedical research on heterogeneous computing environments with appflx,'' \emph{Computational and Structural Biotechnology Journal}, vol.~28, pp. 29--39, 2025.

\bibitem{li2024advances}
Z.~Li, S.~He, Z.~Yang, M.~Ryu, K.~Kim, and R.~Madduri, ``Advances in appfl: A comprehensive and extensible federated learning framework,'' \emph{arXiv preprint arXiv:2409.11585}, 2024.

\bibitem{ryu2022appfl}
M.~Ryu, Y.~Kim, K.~Kim, and R.~K. Madduri, ``Appfl: open-source software framework for privacy-preserving federated learning,'' in \emph{2022 IEEE International Parallel and Distributed Processing Symposium Workshops (IPDPSW)}.\hskip 1em plus 0.5em minus 0.4em\relax IEEE, 2022, pp. 1074--1083.

\bibitem{shrivastava2020dqlearn}
S.~Shrivastava \emph{et~al.}, ``Dqlearn: A toolkit for structured data quality learning,'' in \emph{Proceedings of the IEEE International Conference on Big Data (Big Data)}, 2020, pp. 1644--1653.

\bibitem{gupta2021data}
N.~Gupta, H.~Patel \emph{et~al.}, ``Data quality toolkit: Automatic assessment of data quality and remediation for machine learning datasets,'' \emph{arXiv preprint arXiv:2108.05935}, 2021.

\bibitem{afzal2020data}
S.~Afzal, C.~Rajmohan, M.~Kesarwani, S.~Mehta, and H.~Patel, ``Data readiness report,'' in \emph{Proceedings of the IEEE International Conference on Smart Data Services (SMDS)}, 2020, pp. 42--51.

\bibitem{schelter2018automating}
S.~Schelter, D.~Lange, P.~Schmidt, M.~Celikel, F.~Biessmann, and A.~Grafberger, ``Automating large-scale data quality verification,'' \emph{Proceedings of the VLDB Endowment}, vol.~11, no.~12, pp. 1781--1794, August 2018.

\bibitem{10.14778/3137628.3137631}
\BIBentryALTinterwordspacing
T.~Rekatsinas, X.~Chu, I.~F. Ilyas, and C.~R\'{e}, ``Holoclean: holistic data repairs with probabilistic inference,'' \emph{Proc. VLDB Endow.}, vol.~10, no.~11, p. 1190–1201, Aug. 2017. [Online]. Available: \url{https://doi.org/10.14778/3137628.3137631}
\BIBentrySTDinterwordspacing

\bibitem{ibm_dq_sla_compliance}
{IBM}, ``Data quality sla rule compliance and remediation,'' \url{https://dataplatform.cloud.ibm.com/docs/content/wsj/quality/dq-sla-compliance.html?context=cpdaas&audience=wdp}, 2015, accessed: 2025-04-30.

\bibitem{Jiang_2024}
\BIBentryALTinterwordspacing
X.~Jiang, S.~Sun, J.~Li, J.~Xue, R.~Li, Z.~Wu, G.~Xu, Y.~Wang, and M.~Liu, ``Tackling noisy clients in federated learning with end-to-end label correction,'' in \emph{Proceedings of the 33rd ACM International Conference on Information and Knowledge Management}, ser. CIKM ’24.\hskip 1em plus 0.5em minus 0.4em\relax ACM, Oct. 2024, p. 1015–1026. [Online]. Available: \url{http://dx.doi.org/10.1145/3627673.3679550}
\BIBentrySTDinterwordspacing

\bibitem{Zhang2024FedDQA}
\BIBentryALTinterwordspacing
Z.~Zhang, G.~Chen, Y.~Xu, L.~Huang, C.~Zhang, and S.~Xiao, ``Feddqa: A novel regularization-based deep learning method for data quality assessment in federated learning,'' \emph{Decision Support Systems}, vol. 180, p. 114183, 2024. [Online]. Available: \url{https://doi.org/10.1016/j.dss.2024.114183}
\BIBentrySTDinterwordspacing

\bibitem{rokvic2024liaprivacypreservingdataquality}
\BIBentryALTinterwordspacing
L.~Rokvic, P.~Danassis, S.~P. Karimireddy, and B.~Faltings, ``Lia: Privacy-preserving data quality evaluation in federated learning using a lazy influence approximation,'' 2024. [Online]. Available: \url{https://arxiv.org/abs/2205.11518}
\BIBentrySTDinterwordspacing

\bibitem{du2025dataqualitycontrolfederated}
\BIBentryALTinterwordspacing
Y.~Du, R.~Ye, F.~Yuchi, W.~Zhao, J.~Qu, Y.~Wang, and S.~Chen, ``Data quality control in federated instruction-tuning of large language models,'' 2025. [Online]. Available: \url{https://arxiv.org/abs/2410.11540}
\BIBentrySTDinterwordspacing

\bibitem{eu_industry50}
\BIBentryALTinterwordspacing
{European Commission}, ``Industry 5.0,'' 2021, accessed: 2025-04-14. [Online]. Available: \url{https://research-and-innovation.ec.europa.eu/research-area/industrial-research-and-innovation/industry-50_en}
\BIBentrySTDinterwordspacing

\bibitem{Pearson1901}
K.~Pearson, ``On lines and planes of closest fit to systems of points in space,'' \emph{The London, Edinburgh, and Dublin Philosophical Magazine and Journal of Science}, vol.~2, no.~11, pp. 559--572, 1901.

\bibitem{zhang2023surveyclassimbalancefederated}
\BIBentryALTinterwordspacing
J.~Zhang, C.~Li, J.~Qi, and J.~He, ``A survey on class imbalance in federated learning,'' 2023. [Online]. Available: \url{https://arxiv.org/abs/2303.11673}
\BIBentrySTDinterwordspacing

\bibitem{Kohavi1996}
\BIBentryALTinterwordspacing
R.~Kohavi, ``Adult,'' UCI Machine Learning Repository, 1996, dOI: 10.24432/C5GP7S. [Online]. Available: \url{https://doi.org/10.24432/C5GP7S}
\BIBentrySTDinterwordspacing

\bibitem{fedavgm}
T.-M.~H. Hsu, H.~Qi, and M.~Brown, ``Measuring the effects of non-identical data distribution for federated visual classification,'' \emph{arXiv preprint arXiv:1909.06335}, 2019.

\bibitem{nguyen2022federated}
J.~Nguyen, K.~Malik, H.~Zhan, A.~Yousefpour, M.~Rabbat, M.~Malek, and D.~Huba, ``Federated learning with buffered asynchronous aggregation,'' in \emph{International Conference on Artificial Intelligence and Statistics}.\hskip 1em plus 0.5em minus 0.4em\relax PMLR, 2022, pp. 3581--3607.

\bibitem{li2023fedcompass}
Z.~Li, P.~Chaturvedi, S.~He, H.~Chen, G.~Singh, V.~Kindratenko, E.~A. Huerta, K.~Kim, and R.~Madduri, ``{FedCompass}: efficient cross-silo federated learning on heterogeneous client devices using a computing power aware scheduler,'' \emph{arXiv preprint arXiv:2309.14675}, 2023.

\bibitem{dwork2006differential}
C.~Dwork, ``Differential privacy,'' in \emph{International colloquium on automata, languages, and programming}.\hskip 1em plus 0.5em minus 0.4em\relax Springer, 2006, pp. 1--12.

\bibitem{appflIntegration}
{APPFL Contributors}, ``Data readiness assurance framework in appfl,'' \url{https://appfl.ai/en/latest/tutorials/examples_dr_integration.html}, 2025.

\bibitem{xiao2021experimental}
C.~Xiao and S.~Wang, ``An experimental study of class imbalance in federated learning,'' in \emph{2021 IEEE Symposium Series on Computational Intelligence (SSCI)}.\hskip 1em plus 0.5em minus 0.4em\relax IEEE, 2021.

\bibitem{Chawla2002SMOTE}
N.~V. Chawla, K.~W. Bowyer, L.~O. Hall, and W.~P. Kegelmeyer, ``Smote: Synthetic minority over-sampling technique,'' \emph{Journal of Artificial Intelligence Research}, vol.~16, pp. 321--357, 2002.

\bibitem{Dwork2012Fairness}
C.~Dwork, M.~Hardt, T.~Pitassi, O.~Reingold, and R.~Zemel, ``Fairness through awareness,'' \emph{Proceedings of the 3rd Innovations in Theoretical Computer Science Conference}, pp. 214--226, 2012.

\bibitem{liberty2016stratified}
E.~Liberty, Z.~Karnin, B.~Xiang, L.~Rouesnel, B.~Coskun, R.~Nallapati, J.~Delgado, A.~Sadoughi, Y.~Astashonok, P.~Das \emph{et~al.}, ``Stratified sampling meets machine learning,'' in \emph{Proceedings of The 33rd International Conference on Machine Learning}, ser. Proceedings of Machine Learning Research, vol.~48.\hskip 1em plus 0.5em minus 0.4em\relax PMLR, 2016, pp. 2320--2329.

\bibitem{Tukey1977}
J.~W. Tukey, ``Exploratory data analysis,'' 1977.

\bibitem{Samarati1998Protecting}
P.~Samarati and L.~Sweeney, ``Protecting privacy when disclosing information: k-anonymity and its enforcement through generalization and suppression,'' \emph{Technical report, SRI International}, 1998.

\bibitem{lecun2010mnist}
\BIBentryALTinterwordspacing
Y.~LeCun and C.~Cortes, ``Mnist handwritten digit database,'' 2010. [Online]. Available: \url{http://yann.lecun.com/exdb/mnist/}
\BIBentrySTDinterwordspacing

\bibitem{Krizhevsky09learningmultiple}
A.~Krizhevsky, ``Learning multiple layers of features from tiny images,'' Tech. Rep., 2009.

\bibitem{he2022flamby}
C.~He, S.~Rasouli, I.~Zachariah, P.~Tiwari, P.~Bacon, Y.~Shen, A.~Kotti, O.~Marfoq, H.~Benali, T.~Clozel \emph{et~al.}, ``Flamby: Datasets and benchmarks for cross-silo federated learning in realistic healthcare settings,'' \emph{arXiv preprint arXiv:2210.04620}, 2022.

\bibitem{AI-READI_Consortium_2024}
\BIBentryALTinterwordspacing
A.-R. Consortium, ``Flagship dataset of type 2 diabetes from the ai-readi project (1.0.0),'' 2024. [Online]. Available: \url{https://doi.org/10.60775/fairhub.1}
\BIBentrySTDinterwordspacing

\bibitem{gropp2023delta}
W.~Gropp, T.~Boerner, B.~Bode, and G.~Bauer, ``Delta: Balancing gpu performance with advanced system interfaces,'' National Center for Supercomputing Applications, University of Illinois at Urbana-Champaign, Technical Report, 2023, funded by National Science Foundation (award OAC 2005572).

\bibitem{flamby_ixi_dice}
Owkin, ``{FLamby IXI Dataset},'' \url{https://github.com/owkin/FLamby/blob/main/flamby/datasets/fed_ixi/README.md#prediction-task}, 2021, accessed: 2025-07-14.

\bibitem{Xiao2022Experimental}
C.~Xiao and S.~Wang, ``An experimental study of class imbalance in federated learning,'' \emph{arXiv preprint arXiv:2109.04094}, 2022.

\bibitem{risingwave_federated_learning_2023}
\BIBentryALTinterwordspacing
R.~Labs. (2023) Understanding the impact of class imbalance in federated learning. [Online]. Available: \url{https://risingwave.com/blog/understanding-the-impact-of-class-imbalance-in-federated-learning/}
\BIBentrySTDinterwordspacing

\bibitem{Huang2022Fairness}
W.~Huang, T.~Li, D.~Wang, S.~Du, and J.~Zhang, ``Fairness and accuracy in federated learning,'' \emph{Information Sciences}, vol. 589, pp. 170--185, 2022.

\bibitem{Fisichella2022Partially}
M.~Fisichella, G.~Lax, and A.~Russo, ``Partially-federated learning: A new approach to achieving privacy and effectiveness,'' \emph{Information Sciences}, vol. 610, pp. 1--18, 2022.

\bibitem{Huang2024FedMef}
H.~Huang, W.~Zhuang, C.~Chen, and L.~Lyu, ``Fedmef: Towards memory-efficient federated dynamic pruning,'' in \emph{Proceedings of the IEEE/CVF Conference on Computer Vision and Pattern Recognition (CVPR)}.\hskip 1em plus 0.5em minus 0.4em\relax Computer Vision Foundation, 2024.

\end{thebibliography}


\end{document}